\pdfoutput=1
\documentclass[11pt,a4paper]{article}
\usepackage[T1]{fontenc}
\usepackage[utf8]{inputenc}
\usepackage{amsmath,amssymb}
\usepackage{graphicx}
\usepackage{microtype}
\usepackage[margin=2.5cm]{geometry}
\usepackage{url}
\usepackage[colorlinks=true,allcolors=blue]{hyperref}
\hypersetup{
  pdftitle={Prediction emerges in RNNs trained for perception},
  pdfauthor={Akanksha Gupta, Alejandro Tabas},
  pdfkeywords={predictive processing; Bayesian brain; predictive coding;
               recurrent neural networks; emergent computation; perception}
}
\usepackage{authblk}

\title{Prediction emerges in RNNs trained for perception}

\author[1]{Akanksha Gupta} 
\author[2,3,4,*]{Alejandro Tabas} 

\affil[1]{Aix-Marseille University; Marseille, France}
\affil[2]{Basque Center on Cognition, Brain and Language; San Sebastian, Spain}
\affil[3]{Max Planck Institute for Human Cognitive and Brain Sciences; Leipzig, Germany}
\affil[4]{Ikerbasque, Basque Foundation for Science; Bilbao, Spain}
\affil[*]{tabas@bcbl.eu}

\begin{document}

\maketitle

\section*{Abstract}

The brain is highly proficient at making sense of noisy and ambiguous sensory inputs. Predictive processing hypothesises that this ability relies on prediction. However, it is unclear why the brain would have evolved to predict the sensory world, a computationally expensive process, in order to aid perception. Here we use simulations to argue that prediction naturally emerges in systems optimised for perception. We train recurrent neural networks (RNNs) to denoise a tokenised version of Bach's compositions at a range of noise levels. Afterwards, we enquire whether the states of the networks contain predictive information about the next token. We test this by freezing the RNN weights and training a linear readout from its states on prediction. We compare the performance of the linear readout with that of an independently trained linear benchmark model. The results show that the linear readout from the RNNs outperforms the benchmark model at moderate levels of noise, indicating that the networks rely on a predictive mechanism to support perception. We further show that the responses of the RNNs to sensory inputs are proportional to prediction error. Together, the results demonstrate that neural signatures of predictive processing emerge, without any further training constraints, from optimisation of perception.

\vspace{1em}
\noindent\textbf{Keywords:} predictive processing; Bayesian brain hypothesis;
predictive coding; recurrent neural networks; emergent computation; perception.


\section*{Introduction}

Sensory signals are often noisy and ambiguous. Still, the brain is adept at building a stable representation of the outside world based only on those signals. How the brain performs this mapping remains unclear \cite{Tomic2026, Furutachi2026}. One intriguing possibility, introduced by Helmholtz in the 19th century \cite{Helmholtz1867}, is that the brain encodes a generative model of the environment to support perception. In recent decades, Bayesian frameworks of cognition \cite{Dayan1995, Knill2004, Aitchison2017, Ma2023}, and in particular predictive processing \cite{Rao1999, Friston2003, Clark2013, Keller2018, Millidge2022}, have formalised this idea by proposing that the generative model is used to predict the sensory world, and that those predictions aid in the estimation of the latent causes of the sensory input.

Empirical support for this hypothesis comes mainly from findings reflecting the encoding of the \emph{prediction error}: suppressed responses to expected inputs (e.g.,~\cite{Ulanovsky2003, Nelken2014, Malmierca2015, Parras2017}), mismatch responses to violations of abstract regularities (e.g.,~\cite{Naatanen2007, Paavilainen2013, Tabas2020}), and responses to omitted stimuli (e.g.,~\cite{Bendixen2009, Braga2022, Tabas2025}). Whether these results establish that the brain predicts the sensory input, however, is still a matter of heated debate \cite{May2009, Denham2017, Aitchison2017, Heilbron2018, Westerberg2026}, which centres on why sensory systems, involved solely in processing what is happening now, would dedicate resources to predicting what comes next. A popular argument is that transmitting prediction error is metabolically cheaper than transmitting the sensory inputs \cite{Barlow1961, Manookin2023}. Indeed, previous work has shown that predictive computation emerges from energy constraints in recurrent neural networks \cite{Ali2022, Zhang2025, Nortmann2026}. However, computing prediction errors requires transmitting predictions, which offsets the metabolic gain \cite{Spratling2017, Facchin2025}.

Here, inspired by filtering theory \cite{Kalman1960}, we study whether the ability to predict might have emerged as a direct consequence of the evolutionary pressure to accurately process the sensory inputs; i.e., without any predictive or metabolic constraints, or a presupposed neural architecture. We trained gated RNNs \cite{Cho2014} to denoise sensory inputs, an objective that does not include future states and that we use as a metaphor for perception. We used a tokenised version of Bach's compositions as latent perceptual objects, and created the sensory inputs as noisy observations of those objects. Using Bach's music as inputs guaranteed that the data were generated by a structured generative system, complex enough that predictions are possible and cannot be trivially performed by a first-order Markovian model.

After training in denoising, we enquired whether the hidden states of the RNNs encode predictive information about the upcoming inputs by training a linear readout on prediction. We compared the performance of the linear readout against that of a Markovian benchmark model trained on prediction, and determined that the network encoded predictive information if the readout outperformed the benchmark. Finally, we asked whether the same networks would also display the two neural signatures most often taken as evidence for prediction: responses that scale with prediction error, and states from which prediction error can be read out.

\section*{Results}

  \subsection*{A modelling pipeline to test the emergence of predictions}

    We trained single-layer gated recurrent neural networks (RNNs) to denoise noisy observations $y_t$ of sequences of tokens $x_t$. The tokens were derived from MIDI files of Bach compositions scraped from the internet. Each MIDI file was first transformed into a polyphonic piano-roll matrix. Then, the matrices were tokenised into segments that shared the same active notes; i.e., a new token was emitted only when the set of sounding pitches changed. This resulted in an event-based representation that had no notion of time. The tokens were then collapsed into a 12-dimensional chromatic space to reduce their dimensionality, so that each latent token $x_t \in \{0,1\}^{12}$. We used a total of 700 unique compositions (1,432 MIDI files, as some compositions were split into parts) that were randomly split into training (490 compositions), validation (70 compositions), and testing (140 compositions).

\begin{figure}[tb]
      \centering
      \includegraphics[width=\textwidth]{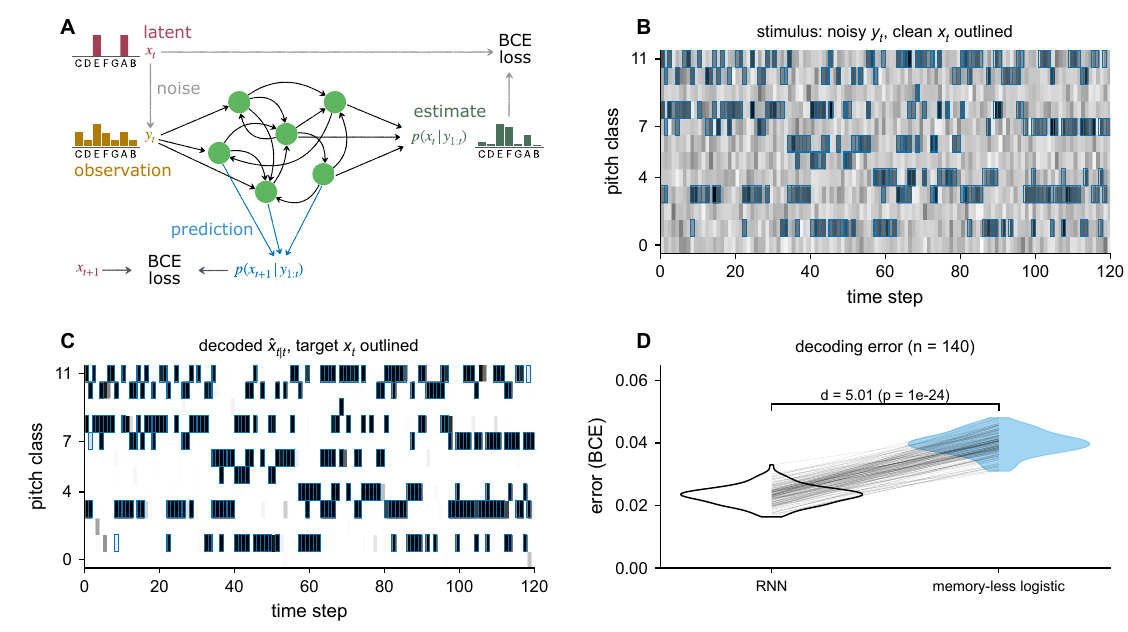}
      \caption{\textbf{Experimental design.}
      \textbf{A)}~Training pipeline. Latent token sequences $x_{1:T}$ are corrupted with Gaussian noise to produce the observations $y_{1:T}$. RNNs are first trained on denoising the observations (training fits the black connections). Afterwards, the remaining weights are frozen and a second linear readout is trained on prediction (training fits the blue connections only).
      \textbf{B)}~The first 120-step excerpt of the stimulus sequence for BWV~111 (test set). The latent tokens $x_t$ are outlined in blue, their corresponding observations $y_t$ ($\sigma=0.25$) are shown in a greyscale heatmap (darker tones for higher values).
      \textbf{C)}~Denoised output of one of the RNNs ($N=64$) for the same excerpt. The network estimates are now shown in greyscale, blue outlines mark the ground-truth latents.
      \textbf{D)}~Decoding error for the same RNN and noise regime measured with binary cross-entropy (BCE) over the 140 test compositions. The RNN denoised the input significantly better than a memory-less logistic regression; thin lines join points from the same composition. Note that BCE is a measure of error: the lower the BCE, the better the performance.}
      \label{fig:design}
    \end{figure}

    Observations $y_t$ were then generated by corrupting the latents with Gaussian noise $y_t = x_t + \sigma\varepsilon_t$ with $\varepsilon_t \sim \mathcal{N}\left(0, \mathbf{I}\right)$ (Figure~\ref{fig:design}A and~B). We used 14 noise amplitudes, ranging from essentially noiseless ($\sigma = 0.001$) to a signal-to-noise ratio of 0.5 ($\sigma = 2$), to measure a potential dependence of the emergence of prediction on the noise regime.

    RNNs were first trained on denoising (a metaphor for perception) with a binary cross-entropy (BCE) loss. RNNs received as inputs sequences of observations $y_{1:T}$ and were trained to recover, through a linear readout, the latent tokens $x_{1:T}$ (Figure~\ref{fig:design}A). After training was completed, we froze the weights of the RNNs and trained a second linear readout on prediction; i.e., the RNNs were presented with the same observations $y_{1:T-1}$, but the linear readout was trained on predicting, from the hidden states at $t$, the subsequent latent $x_{t+1}$. We trained RNNs of different sizes ($N \in \{8, 16, 32, 64, 128, 256\}$ hidden units) to measure a potential dependence of the emergence of prediction on the capacity of the network.

    In addition to the RNNs, we fitted five benchmark models to contextualise the performance of the target RNNs. First, to bound the denoising problem, we fitted a memory-less logistic regression mapping a single observation $y_t$ onto the concurrent latent token $x_t$. The remaining four benchmark models bound the prediction problem: (1)~a marginal predictor that systematically predicts the global prior independently of the input, quantifying the optimal chance level under BCE; (2)~a memory-less logistic regression predicting the next token $x_{t+1}$ from a single observation $y_t$, i.e., a first-order Markovian analogue of the RNNs' readout without access to the hidden states, which we term the \emph{liberal} Markovian benchmark because it sees exactly the same input as the RNNs; (3)~an alternative version of (2) where the model was fed $x_t$ instead of the noisy observation, analogous to a linear readout of the RNNs on a perfectly denoised input, which we term the \emph{conservative} Markovian benchmark; (4)~an RNN of the same class and size as the largest networks of the sweep ($N = 256$), trained end-to-end on prediction, quantifying the best prediction performance achievable within this model class from noisy inputs.

    We expected that an RNN in which prediction emerges would show a prediction BCE below that of the conservative Markovian benchmark~(3), and that an RNN in which prediction does not emerge would show a prediction BCE between those of the conservative~(3) and the liberal~(2) Markovian benchmarks. In neither case did we expect the RNN to reach benchmark~(4), the prediction-trained RNN, which we take as the empirical performance ceiling.

  \subsection*{RNNs outperform memory-less models at perception}

    RNNs successfully learned to denoise the observations (Figure~\ref{fig:design}C; Supplementary Figure~\ref{figS:decoding}). In the 64-unit RNN trained on $\sigma = 0.25$ (case-example used throughout the manuscript), the network (mean $\mathrm{BCE} = 0.023$) outperformed the memory-less logistic regression (mean $\mathrm{BCE} = 0.040$; Figure~\ref{fig:design}D; Cohen's $d = 5.01$, $p < 10^{-20}$), indicating that the network exploited its memory trace during denoising.

    RNNs systematically outperformed the memory-less logistic regression benchmark at network sizes $N \geq 16$ across noise levels (Supplementary Figure~\ref{figS:decoding}; all $p < 10^{-22}$, with effect sizes ranging from $d = 1.03$ to $d = 17.53$). The 8-unit networks were, however, outperformed by the logistic regression for $0.15 \leq \sigma \leq 0.2$, as expected from a hidden state narrower than the $12$-dimensional observation it must represent. For $N \geq 16$, the advantage in raw BCE was negligible in noise regimes $\sigma \leq 0.1$, increased steadily with the noise level up to $\sigma = 0.8$, and monotonically decreased thereafter. At $\sigma = 0.8$, the denoising advantage increased monotonically with $N$ up to $N = 128$, and declined slightly at $N = 256$.

    These results are consistent with RNNs exploiting memory to improve their denoising estimates. Although, strictly speaking, RNNs can approximate non-linear mappings that could theoretically outperform the memory-less logistic regression, the growth of the advantage with $\sigma$ up to $\sigma = 0.8$ favours the memory interpretation.

  \subsection*{Predictive information emerges without a predictive objective}

\begin{figure}[bt]
      \centering
      \includegraphics[width=\textwidth]{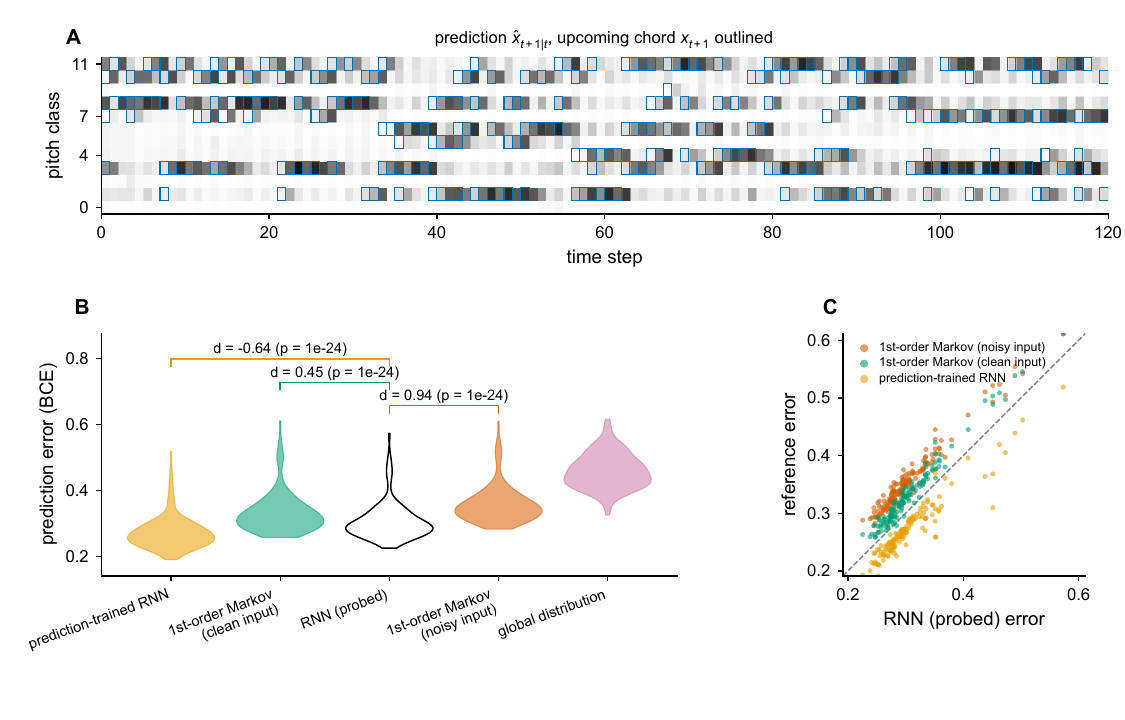}
      \caption{\textbf{Predictive information emerges in an RNN trained on perception.}
      \textbf{A)}~Case-example RNN ($\sigma = 0.25$, $N = 64$) predictions (greyscale heatmap) for the same excerpt shown in Figure~\ref{fig:design}; ground-truth latents $x_{t+1}$ are outlined in blue.
      \textbf{B)}~The case-example RNN (white) shows a test-set prediction BCE consistent with the internal encoding of predictions. The network BCE is compared against four benchmark models (from left to right): (yellow) RNN fully trained on prediction, the empirical performance ceiling; (green) conservative Markovian benchmark, a first-order model predicting $x_{t+1}$ from the ground-truth $x_t$, a conservative threshold for a model without predictive information; (orange) liberal Markovian benchmark, a first-order model predicting $x_{t+1}$ from the same input as the RNN, $y_t$, a liberal threshold for a model without predictive information; and (pink) global predictive distribution, the optimal chance-level model under BCE.
      \textbf{C)}~Performance is consistent across compositions. The BCE of the case-example RNN ($x$-axis) is compared against the BCE of the best three benchmark models ($y$-axis) for each composition. Points above the diagonal are compositions in which the RNN outperforms the benchmark.}
      \label{fig:predictions}
    \end{figure}

    The predictive readout of the case-example network ($N = 64$, $\sigma = 0.25$) outperformed both Markovian benchmark models (Figure~\ref{fig:predictions}; Cohen's $d = 0.45$, $p < 10^{-20}$ in the comparison with the conservative Markovian benchmark, which uses the ground-truth $x_t$ as input). Therefore, at least in this noise regime, the denoising-trained RNNs encoded predictive information about the upcoming token that was not directly deducible from the previous denoised input. As expected, however, the network did not reach the empirical performance ceiling and was outperformed by the prediction-trained RNN ($d = -0.64$, $p < 10^{-20}$). The results were consistent across the compositions of the corpus (Figure~\ref{fig:predictions}C).

  \subsection*{Predictive information emerges only at intermediate noise regimes}

\begin{figure}[htb]
      \centering
      \includegraphics[width=\textwidth]{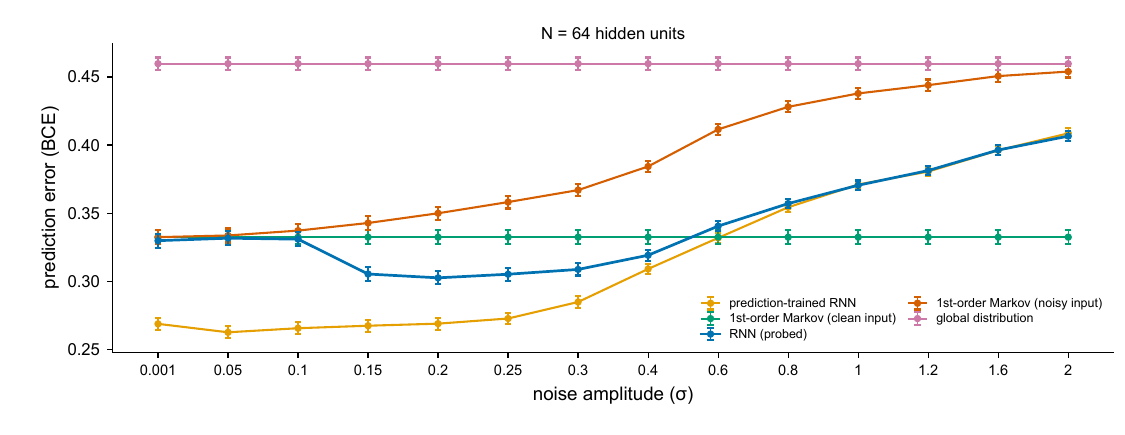}
      \caption{\textbf{Predictive information emerges at moderate noise levels.}
      Prediction BCE for the 64-unit denoising-trained RNN (blue) in comparison to the benchmark models for different levels of noise ($x$-axis). At $\sigma \leq 0.1$, the RNNs show similar performance to the two Markovian models (orange and green), consistent with a lack of predictive information. At $0.15 \leq \sigma \leq 0.4$, the RNN systematically outperforms both Markovian models, demonstrating the encoding of predictive information. At $\sigma = 0.6$, the performance of the denoising-trained RNN, the prediction-trained RNN (yellow; empirical performance ceiling), and the conservative Markovian benchmark (green) merge. At $\sigma > 0.6$, both RNNs are outperformed by the conservative Markovian benchmark, which receives as inputs the noiseless latents $x_t$ and thus presents the same performance at all noise levels. Neither of the RNNs, however, is outperformed by the liberal Markovian benchmark (orange), which predicts the next token based on the noisy observations $y_t$, and asymptotically approaches the performance of the chance-level model (pink) at $\sigma=2$. This pattern of results demonstrates that the denoising-trained RNN encodes predictive information at noise regimes $0.15 \leq \sigma \leq 0.4$.}
      \label{fig:noiseregime}
    \end{figure}

\begin{figure}[p]
      \centering
      \includegraphics[width=\textwidth]{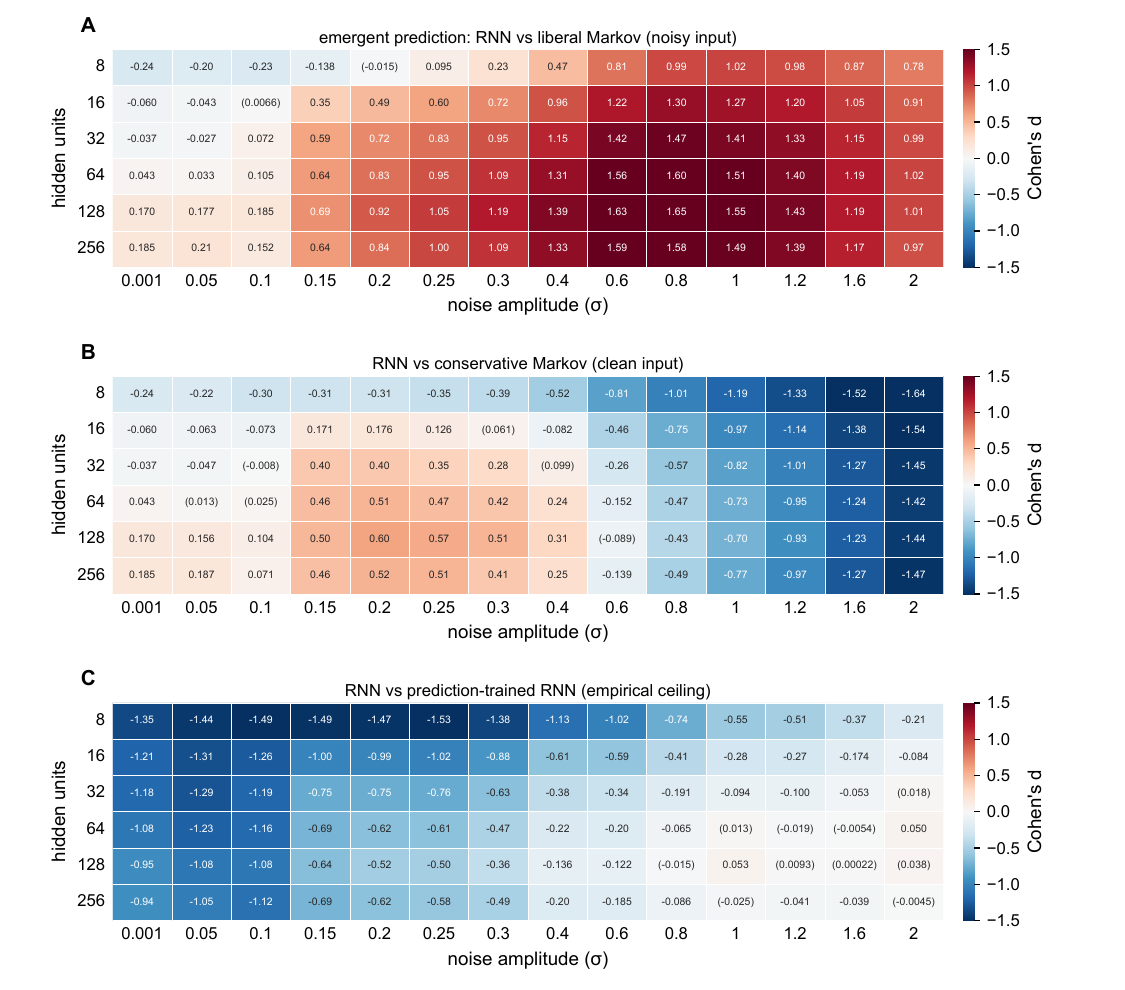}
      \caption{\textbf{Emergent prediction across noise amplitude and network size.}
      Cohen's $d$ for the contrast between each benchmark and the denoising-trained RNN (benchmark
      minus RNN, so warm colours mark the network's advantage) across the 14 noise amplitudes
      ($x$-axis) and the 6 network sizes ($y$-axis), computed over the 140 test compositions after
      averaging each composition across the five runs.
      \textbf{A)}~Against the liberal Markovian benchmark, which receives the same noisy
      observations $y_t$ as the RNN.
      \textbf{B)}~Against the conservative Markovian benchmark, which receives the noiseless
      latents $x_t$.
      \textbf{C)}~Against the prediction-trained RNN, the empirical performance ceiling the probed
      readout is not expected to reach.
      Values enclosed in parentheses mark cells that do not survive Bonferroni correction over the
      84 cells of the design.}
      \label{fig:sweep}
    \end{figure}

    The emergence of predictive information depended on the noise level under which the RNNs were trained (Figure~\ref{fig:noiseregime}). The prediction performance of the 64-unit network (blue line) remained relatively stable for $\sigma \leq 0.1$ but improved abruptly at $\sigma = 0.15$, despite operating on noisier inputs $y_t$. As the noise level increased further, the performance of the RNN started to decline. The denoising-trained RNN performed comparably to the conservative Markovian benchmark (green line) for $\sigma \leq 0.1$, outperformed it for $0.15 \leq \sigma \leq 0.4$, and fell behind it thereafter. However, this decline was mirrored by the prediction-trained RNN (yellow), whose performance converged with that of the denoising-trained RNN for $\sigma \geq 0.6$. Crucially, the denoising-trained RNN performed systematically better than the liberal Markovian benchmark, which receives the exact same inputs (orange line), for $\sigma \geq 0.1$. Together, these results indicate that predictive information emerges in the denoising-trained RNNs at some point between $\sigma = 0.1$ and $\sigma = 0.15$. This result parallels the pattern observed in the denoising advantage of the RNNs over the memory-less regression, which becomes substantial between $\sigma = 0.15$ and $\sigma = 0.25$ (Supplementary Figure~\ref{figS:decoding}). Results were consistent across RNNs with $N \geq 32$ (Figure~\ref{fig:sweep}).

  \subsection*{Activation elicited by the sensory input scales with prediction error}

    We next enquired whether the networks, if they were implemented as neural populations in the brain, would elicit responses (e.g., event-related potentials) proportional to prediction error. We measured prediction error as $e_t = x_{t} - \hat{x}_{t|t-1}$, where $\hat{x}_{t|t-1}$ is the output of the linear readout trained on prediction over the denoising-trained RNN. We assumed responses would be proportional to the magnitude of the change in activation of the RNN elicited by the new observation, $\Delta h_t = h_{t} - h_{t-1}$, where $h_t$ are the activations of the hidden units after having received $y_t$. All correlations below were computed between the squared magnitudes $|e_t|^2$ and $|\Delta h_t|^2$ across the time steps of a composition.

\begin{figure}[p]
      \centering
      \includegraphics[width=\textwidth]{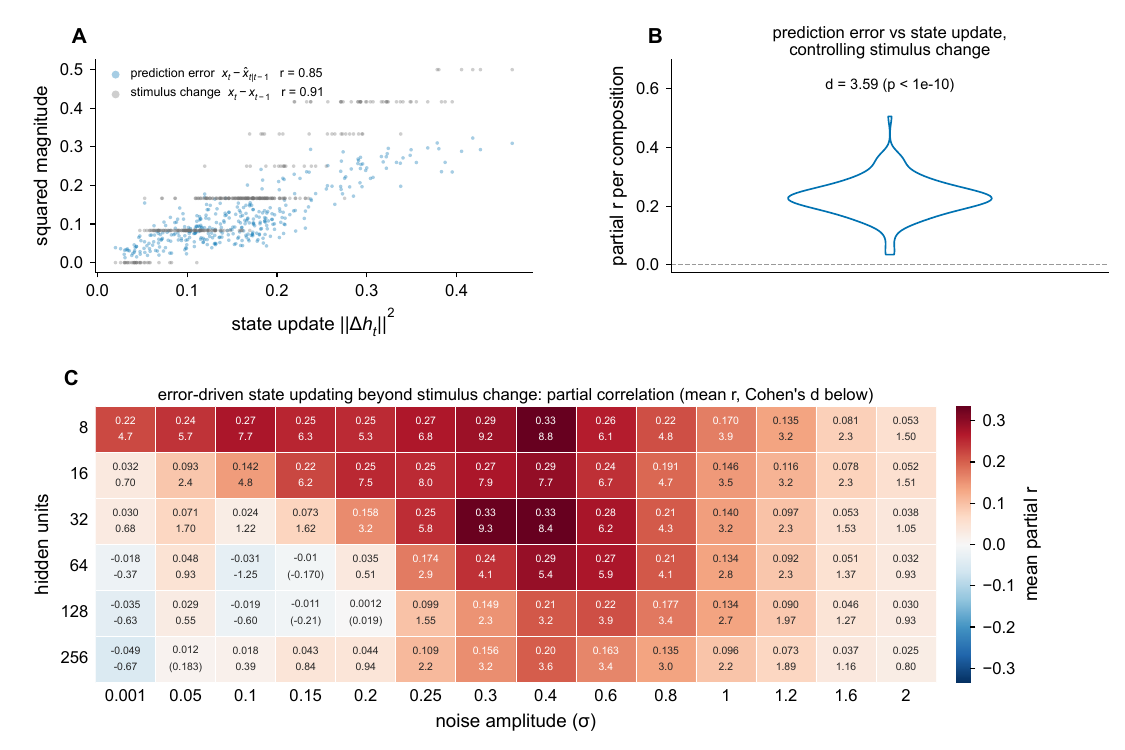}
      \caption{\textbf{State updates correlate with prediction error.}
      \textbf{A)}~Correlation between state updates and prediction error for the case-example network ($\sigma = 0.25$, $N = 64$) for a single composition (BWV~111). Although state updates are strongly correlated with prediction error (blue), they are also correlated with the update of the latent $x_t-x_{t-1}$ underlying the observations (grey; note that the latent update is discrete because the latents are binary).
      \textbf{B)}~Partial correlation between prediction error and state update controlling for the latent update across all compositions of the test set for the case-example network.
      \textbf{C)}~Mean partial correlation and effect sizes across the test set for all network sizes ($y$-axis) and noise regimes ($x$-axis). Values of Cohen's $d$ enclosed in parentheses indicate contrasts that were not significant against the null hypothesis $d = 0$ after Bonferroni correction for 84 comparisons.}
      \label{fig:pe}
    \end{figure}

    The update of the case-example network ($N = 64$, $\sigma = 0.25$) closely matched prediction error in the example composition (Figure~\ref{fig:pe}A, blue), with a correlation $\rho = 0.85$, $p < 10^{-20}$. However, these correlations may be partially caused by the changes in the latent $x_t$, $|x_t - x_{t-1}|^2$, which also correlate with state updates (Figure~\ref{fig:pe}A, grey). We therefore ran a partial correlation between state update and prediction error, controlled for the change in the latent. Partial correlations were strictly above zero for all the compositions in the test set (Figure~\ref{fig:pe}B; Cohen's $d = 3.59$, $p < 10^{-20}$). This result demonstrates that the RNN dynamics would have yielded neural responses that could be interpreted as prediction error.

    This was not the case in every noise regime. In the 64-unit networks the partial correlation rises sharply at $\sigma = 0.25$, peaks at $\sigma = 0.4$ ($\rho = 0.29$), and decays thereafter, falling below $0.1$ by $\sigma = 1.2$ (Figure~\ref{fig:pe}C). This pattern is expected from the previous results. At low noise levels, the observation can simply be passed forward by the networks, and the state update merely reflects this pass-forward computation. On the other hand, when the observation is dominated by noise, the network update tracks neither the stimulus nor the prediction error. It is at intermediate noise levels, at which the networks leverage their predictions, that state updates reflect prediction error.

    A similar profile is apparent at other network sizes $N$, although the noise regime that promotes the encoding of prediction error widens as network size $N$ increases. This effect is probably driven by the dependence of the control term of the partial correlation on network capacity: all networks with $N \geq 16$ solve the denoising task about equally well (test BCE $0.10$--$0.12$ at $\sigma = 0.4$), but they do not use their units the same way. A network with a larger state space can afford to carry a redundant, stimulus-locked copy of the current latent token alongside whatever temporal integration the noise demands. The 8-unit networks, with a capacity lower than the dimensionality of the inputs, show a pattern of results difficult to interpret, with substantial partial correlations at all noise levels. A possible interpretation of these results is that the correlations reflect a $12 \rightarrow 8$ compression strategy of the networks that is not successfully corrected by a linear control.

  \subsection*{The networks that encode predictions also encode prediction error}

    Next, we enquired whether the network states \emph{encode} prediction error; i.e., whether the prediction error vector $e_t$ could be decoded through a linear readout of the hidden states. This is a stronger condition than the state update magnitude merely correlating with prediction error.

    To do this, we fitted ridge regressions from the hidden state $h_t$ to the prediction error $e_t = x_t - \hat{x}_{t|t-1}$ using the validation set. We then evaluated the regressions using the 140 compositions of the test set. We contextualised the decoding accuracy using three benchmark models also trained on the validation set: (1)~a constant decoder emitting the expected prediction error vector; (2)~a decoder predicting $e_t$ from the current observation $y_t$; and (3)~a decoder predicting $e_t$ from the current and past observations $y_t$ and $y_{t-1}$. Since the RNN states may encode information about the current and previous observations, we determined that the RNNs encoded prediction error only if they outperformed the third and most stringent of the benchmarks.

\begin{figure}[p]
      \centering
      \includegraphics[width=\textwidth]{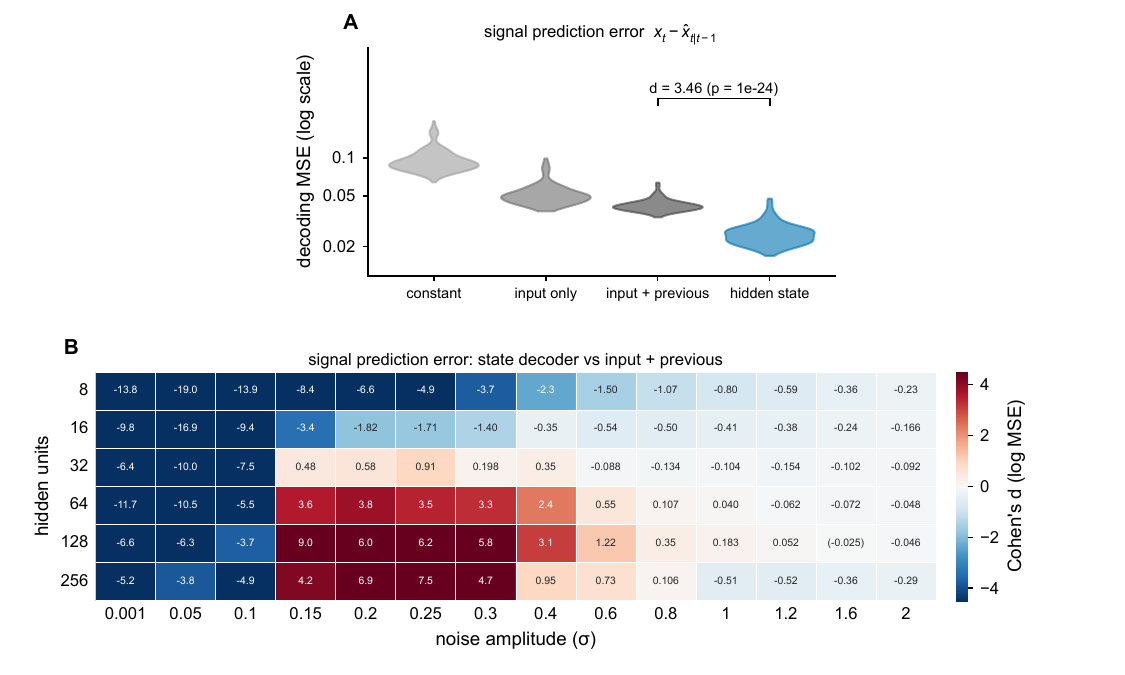}
      \caption{\textbf{RNNs trained at moderate noise levels encode prediction error.}
      \textbf{A)}~Mean squared error (MSE; in log scale) of the prediction error decoded from the RNN hidden states for the case-example RNN ($\sigma = 0.25$, $N = 64$; blue) in comparison with three benchmark models: a constant (dark grey), a linear regressor using as input the current observation (grey), and a linear regressor using as input the current and past observations (light grey). The linear readout outperforms all the benchmarks in all compositions.
      \textbf{B)}~Cohen's $d$ for the contrast between the RNN readout and the best-informed benchmark, which takes as input the current and past observations, across network sizes ($y$-axis) and noise levels ($x$-axis). Values of Cohen's $d$ enclosed in parentheses mark results that were not significant after Bonferroni correction for 84 comparisons.}
      \label{fig:decoding}
    \end{figure}

    The linear readout of the case-example network outperformed all the benchmarks in all compositions (Figure~\ref{fig:decoding}A; $d = 3.46$, $p < 10^{-20}$ for the comparison against the most stringent benchmark), demonstrating that it encodes prediction error. Robust encoding of prediction error, however, is only apparent for 64-unit networks trained on noise regimes with $0.15 \leq \sigma < 0.6$; i.e., the noise regimes for which the RNNs outperformed the Markovian benchmarks (Figure~\ref{fig:decoding}B). This pattern is replicated for larger RNNs, but prediction error encoding is not apparent in networks with $N < 32$ units.

\section*{Discussion}

  Predictive processing proposes that the brain continuously predicts the sensory input to aid perceptual processing \cite{DeLange2018, Furutachi2026}. Whether this is indeed the case is still under discussion. Some authors argue that prediction, which might itself be flawed, is not necessary or desirable for perception \cite{Huettig2015}; others have argued that predictions are only useful under restricted regimes of noise and channel capacity \cite{Chalk2017, Aitchison2017}. Here, we trained RNNs on a perceptual task without predictive objectives, and enquired whether the basic components of predictive processing emerged from training. Networks trained on intermediate noise regimes acquired the three computational signatures of predictive processing: their hidden states encoded predictions about the upcoming inputs, their state updates correlated with prediction error, and their states encoded prediction error. On the other hand, networks trained on low- or high-noise regimes did not develop a demonstrable capacity to predict the sensory input. These results clarify why a perceptual system would have evolved to predict merely in order to perceive, and offer a testable qualitative hypothesis on the perceptual regimes where predictive processing is expected to have emerged.

  Previous work showed that prediction emerges in RNNs trained on perception under energy constraints \cite{Ali2022, Zhang2025}. These works echo a line of argument within the predictive processing community holding that the transmission of prediction error (what cannot be accounted for by the predictions) is, on the long-term average, metabolically more efficient than the transmission of the sensory inputs \cite{Barlow1961, Manookin2023}. However, to compute prediction errors, neural systems need to constantly transmit predictions, which renders the energy gain negligible \cite{Spratling2017, Facchin2025}. Even if RNNs whose activations are constrained autonomously develop the capacity to predict, it is not clear how well the energy savings would extrapolate to perceptual problems that require long-range connections or the prediction of complex hierarchical structures.

  Here we show, for the first time, that whether a system implements predictive processing may not be conditioned on the architecture or internal constraints of the system, but on the requirements of the task. Specifically, we show that predictive computation emerges at intermediate noise levels. This result is consistent with filtering theory \cite{Kalman1960}: when the communication channel is nearly noiseless, passive decoding is sufficient to recover the latents and encoding a generative model of the inputs does not increase the accuracy of the system. On the other hand, in communication channels dominated by noise, the observations are not informative about the latents and optimal estimates drift towards the marginal statistics. Prediction is worth computing only in between these two regimes.

  The emergence thresholds observed in our networks can be understood using detection theory. Predictions emerged reliably in the networks trained in regimes where the Gaussian noise had standard deviations $0.15 \lesssim \sigma \lesssim 0.4$ relative to the signal amplitude ($x_t \in \{0,1\}^{12}$). Assuming that each of the 12 dimensions $k$ of the latent $x_t$ is sampled $x_t^k \overset{\text{iid}}{\sim} \mathrm{Bern}(0.5)$, the probability of recovering the latent exactly from a single observation $y_t \sim \mathcal{N}(x_t, \sigma^2 \mathbf{I})$ is $P(\text{correct} \mid \sigma) = \Phi\!\left( \frac{1}{2\sigma} \right)^{\!12}$ \cite{Green1966, Macmillan2021}, where $\Phi$ is the cumulative distribution function of the standard Gaussian. At the lower bound, $P(\text{correct} \mid \sigma < 0.15) > 0.99$: the instantaneous observation is already almost always sufficient. Above this point, recovery degrades steeply and predictions become the only way to aid accurate reconstruction. At the upper bound, $P(\text{correct} \mid \sigma > 0.4) < 0.27$, single observations become individually uninformative, and predictions fail to emerge because the observations no longer support the learning of the statistical structure they would need to exploit.

  This pattern establishes a qualitative hypothesis about the domains in which predictive processing may apply if its emergence is grounded only in the demands of the perceptual task, rather than in external factors like energy optimisation. If that were the case, we would expect predictions to be encoded only in neural populations that are shaped to operate under moderate noise levels. This does not necessarily mean that predictions emerge only under noisy sensory inputs: the classical paradigms typically used to demonstrate the encoding of prediction error, such as the oddball paradigm, use sensory inputs with near-absent noise levels \cite{Naatanen2001, Garrido2009, Heilbron2018, Keller2018, Walsh2020, Tabas2021}. However, sensory processing units optimised for processing under moderate levels of noise might have developed the ability to predict, even if predictions are not directly employed during perception. On the other hand, sensory processing units that have developed to process noiseless data should not be able to predict the sensory input, even when tested using noisy observations.

  The expression of predictive processing might not only vary across sensory systems that developed to process inputs with different noise regimes, but also across individuals who grew up in environments that contained higher or lower sensory noise. In the extreme, these noise regimes might underlie neuropathological conditions related to an over-reliance (e.g., psychosis \cite{Keller2024}) or under-reliance (e.g., developmental dyslexia \cite{Jarvikyla2026} or autism \cite{VanSchalkwyk2017}) on predictions. The present results provide predictive processing with a normative ground based solely on perceptual accuracy, which could be key to identifying the factors responsible for predictive processing and its dysfunction.

\section*{Methods}
  \subsection*{Data}

    MIDI files were scraped from seven public archives\footnote{jsbach.net, metronimo.com, piano-midi.de, sound.jp/bach-ken, dardel.info, classicalarchives.com, and suzumidi.com.} and matched to their BWV catalogue number; duplicates across archives were removed. Files sharing a BWV number (different movements or arrangements of the same work) were concatenated into a single composition, separated by 20 silent tokens. The 700 compositions were split by BWV number rather than by file, so that no movement of a test composition was ever seen during training.

    Noise realisations were drawn independently at every presentation of a training chunk, which makes the effective training set unbounded and prevents the networks from memorising a particular noise combination. In contrast, the noise realisations used to generate observations for the validation and test sets were used consistently during the evaluation of the models and benchmarks, using seeds based on their BWV number.

  \subsection*{Target networks}

    The recurrent networks were single-layer gated recurrent unit (GRU) networks~\cite{Cho2014} with the hidden state initialised to zero at the beginning of each composition. Both readouts were affine maps followed by an element-wise sigmoid, and their outputs were clamped to $[10^{-6}, 1-10^{-6}]$. Since targets were compositional (i.e., a single target $x_t$ usually had more than one active element), outputs were treated as $12$ independent Bernoulli probabilities rather than as a distribution over chords and scored with the mean binary cross-entropy (BCE) over dimensions and time steps.

    Both training stages used Adam~\cite{Kingma2015} with a learning rate of $0.02$ on batches of $512$ sequences of $512$ tokens: $20{,}000$ batches for denoising and $3{,}000$ for the prediction readout (Supplementary Figure~\ref{figS:loss}). In the first training stage the prediction head was frozen at its random initialisation and did not contribute to the gradient; in the second stage, the recurrent weights and the denoising head were frozen and only the prediction head was fitted, so that the hidden dynamics of the RNN were identical to those learned during the denoising objective. The batch budget was fixed so that every cell of the design received the same amount of training. However, validation error was evaluated every $100$ batches and the parameters were restored to the best-scoring checkpoint at the end of each stage. Every combination of noise amplitude and network size was trained five times from independent initialisations, resulting in a total of $14 \times 6 \times 5 = 420$ networks. Unless otherwise specified, results reported for a single network correspond to the first of the five trainings, and results reported across the design are averaged over the five trainings within each composition before aggregation.

  \subsection*{Benchmark models}

    The two Markovian benchmarks and the memory-less denoising benchmark were fitted as $12$ independent logistic regressions, one for each of the 12 dimensions (pitch classes), on the training compositions. We used negligible regularisation ($C = 10^{6}$). Dimensions that were never active in the training set were assigned their clamped marginal frequency instead of a fitted regression.

    The marginal predictor uses the per-dimension frequency over the training set (a constant vector) as prediction, computed by averaging each composition's frequency so that compositions rather than tokens are weighted equally. This is the optimal constant-vector prediction in the sense that it minimises binary cross-entropy.

    The prediction-trained network was a GRU with $N = 256$, trained end-to-end on the prediction objective with the same learning rate, batch size, and chunk size as the target networks, over a single stage of $25{,}000$ batches.

  \subsection*{Partial correlations}

    All the per-timestep series were aligned on the predicted token: index $t$ refers to the token whose prediction $\hat{x}_{t|t-1}$ was formed from the state $h_{t-1}$. Squared magnitudes were averaged over their components, over the 12 dimensions for the error terms and over the $N$ units for the state update, so that they are comparable across network sizes.

    Partial correlations between prediction error $e$ and state update $\Delta$ given the change in the latent tokens $\delta$ were computed as $\rho_{e, \Delta \mid \delta} = \frac{\rho_{e,\Delta} - \rho_{e,\delta} \, \rho_{\Delta, \delta}}{\sqrt{(1 - \rho_{e,\delta}^{2})(1 - \rho_{\Delta, \delta}^{2})}}$, where $\rho_{x,y}$ is Pearson's correlation between $x$ and $y$. Since the latents are binary, $\delta$ takes only the twelve values $k/12$ given by the number of dimensions that changed, and the expression above therefore removes the linear component of a discrete control. Replacing it with indicator variables for each level, with separate onset and offset counts, or with the full 12-dimensional change vector altered $\rho_{e, \Delta \mid \delta}$ by less than $0.02$ at every noise amplitude tested.

    Silent tokens were excluded from all the prediction-error analyses, since a state that has been idling through silence otherwise couples the series spuriously.

  \subsection*{Prediction error decoders}

    The prediction-error decoders were ridge regressions fitted on the validation set. Penalties were selected per output dimension by leave-one-out cross-validation over seven values spanning $10^{-3}$ to $10^{3}$.

    The three benchmark decoders used to contextualise the performance of the RNN readouts were fitted on the same data with the same procedure and differ only in their inputs.

  \subsection*{Statistical analyses}

    The composition was the unit of analysis in every test ($n = 140$). Significance was assessed using Wilcoxon signed-rank tests, and effect sizes were reported as Cohen's $d$. Values reported in heatmaps across all combinations of network size and noise levels were Bonferroni-corrected over the $84$ cells of the design.

    Decoding errors were compared on a logarithmic scale. The binary cross-entropies of the prediction and denoising analyses are bounded and well behaved, and were compared on their raw scale.

  \subsection*{Implementation}

    Models were implemented in PyTorch~\cite{Paszke2019}. MIDI parsing was performed with mido. Benchmark models and statistical analyses used scikit-learn~\cite{Pedregosa2011} and SciPy~\cite{Virtanen2020}.

    The corpus retrieval, training, benchmarking, and analysis code, together with the fitted weights and the analysis outputs, is available at \url{https://github.com/qtabs/BayesPlusBach}.

\section*{Data and code availability}
All code required to run the simulations and produce the results of the paper is available at \url{https://github.com/qtabs/BayesPlusBach}.

\section*{Acknowledgements}
The authors would like to thank BrainHack Donostia 2024 and the team that kickstarted this project during the hackathon (in alphabetical order): Giada Antonicelli, Ekain Arrieta, Emmanuele Ciardo, David Hern\'andez-Guti\'errez, Ana Joya, Cl\'ementine L\'evy-Fidel, Ihintza Malharin, Nirmitee Mulay, Christoforos Souganidis, and Antje Walter.

AT is funded by the ERC (SynPrePro, 101115798) and the Spanish AEI (RyC-2022-036078-I; PID2025-169017NA-I00).

\bibliographystyle{IEEEtran}
\bibliography{bib}

\clearpage
\appendix

\section*{Supplementary Materials}

\renewcommand{\thefigure}{S\arabic{figure}}
\renewcommand{\theHfigure}{S\arabic{figure}} 
\setcounter{figure}{0}

\begin{figure}[htbp]
  \centering
  \includegraphics[width=\textwidth]{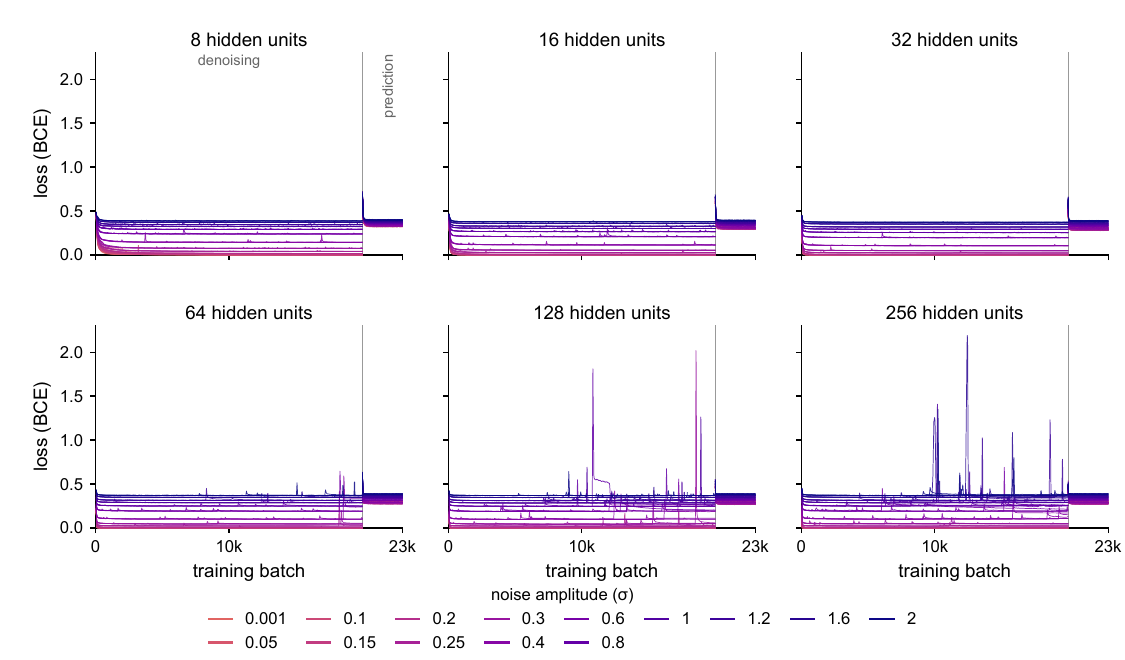}
  \caption{\textbf{Training loss of the target RNNs.} Each panel shows the training loss (BCE; $y$-axis) across noise levels (colour coded). Each model was trained five times independently (runs), and the training losses across runs are superimposed as thin lines. The $x$-axis runs across the batch number: the first 20k batches correspond to the denoising loss during the fitting of the RNN, the last 3k batches correspond to the prediction loss during the fitting of the prediction linear readout. Each panel displays the loss of networks with a different number of hidden units.}
  \label{figS:loss}
\end{figure}

\begin{figure}[htbp]
  \centering
  \includegraphics[width=\textwidth]{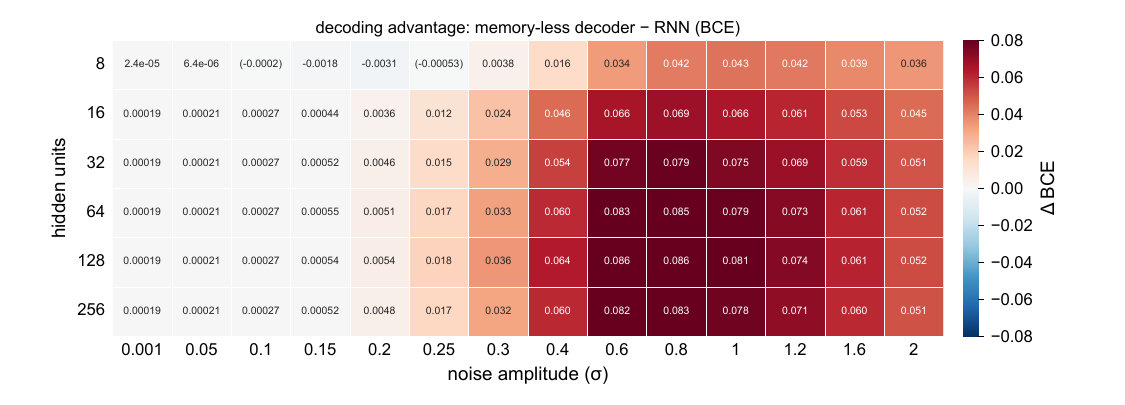}
  \caption{\textbf{Denoising advantage across networks.} Difference in denoising error (BCE) between the memory-less logistic decoder and the RNNs. Cells where the BCE difference is enclosed in parentheses mark results that were not significant after Bonferroni correction for 84 comparisons.}
  \label{figS:decoding}
\end{figure}

\end{document}